# Detection of Equipment Faults Before Beam Loss

*J. Galambos*
ORNL, Oak Ridge, TN, USA

**Abstract**
High-power hadron accelerators have strict limits on fractional beam loss. In principle, once a high-quality beam is set up in an acceptable state, beam loss should remain steady. However, in practice, there are many trips in operational machines, owing to excessive beam loss. This paper deals with monitoring equipment health to identify precursor signals that indicate an issue with equipment that will lead to unacceptable beam loss. To this end, a variety of equipment and beam signal measurements are described. In particular, several operational examples from the Spallation Neutron Source (SNS) of deteriorating equipment functionality leading to beam loss are reported.

## 1   Introduction

Prevention of beam loss is a primary concern for high-power, high-intensity proton machines, to avoid instantaneous damage and longer-term residual activation build-up. The typical rule of thumb for avoiding residual activation build-up is to maintain beam loss below 1 W/m (for beam energies above ~100 MeV) [1]. For megawatt-level beams, this corresponds to a small fractional loss ($10^{-6}$), which can be a challenge to measure, much less anticipate. Indeed, direct beam loss measurements are often the first sign of a developing equipment issue. Sudden and catastrophic equipment failures are easy to detect and diagnose, and result in direct shut-down of the beam. The more challenging task is to detect the slow gradual loss of equipment performance leading to very small impacts on beam transport, yet significant enough to increase the beam loss above the 1 W/m level. Detecting these slow and exceedingly slight equipment degradations is the subject of this paper.

Direct monitoring of the equipment directly related to beam transport is a straightforward method of anticipating issues that can lead to beam loss. This monitoring involves system measurements of magnets, power supplies, RF systems, vacuum, sources, and rotating equipment (pumps). Quantities that are monitored include temperature, voltage, and current. Examples of changes in these quantities that affect beam loss will be outlined in Section 3.

We note that the most sensitive measure of changes in the transport of high-power hadron beams is often beam loss measurement. It is often possible to continue running, even with a modest increase in beam loss (say from 0.1 to 0.2 W/m). Careful attention to changes in beam loss levels, even if they are at acceptable levels, is a valuable method of detecting incipient equipment degradation. The beam itself is a quite useful probe of equipment health. Finally, other beam measurements can be useful indictors of equipment health, as discussed in Section 4.

Most of the examples for beam loss and equipment issues are taken from the Spallation Neutron Source (SNS), which is a neutron scattering facility. It includes a high-power 1.4 MW proton accelerator [2].

## 2 Preparing for the beam

### 2.1 Reaction time-scales

Before discussing equipment failure precursors, it is useful to understand the reaction time-scales needed to protect equipment from gross beam loss. Figure 1 shows the time required for copper to increase in temperature by 100°C when subject to a 1 mm² cross-section proton beam of varying average current and energy. This can be thought of as a characteristic response time to protect the machine from catastrophic damage from the beam (or approaching the 'melting metal' stage). The reaction time is shorter at low energies, owing to the shorter beam penetration length. For average currents above 10 mA, the reaction time is generally less than 1 ms.

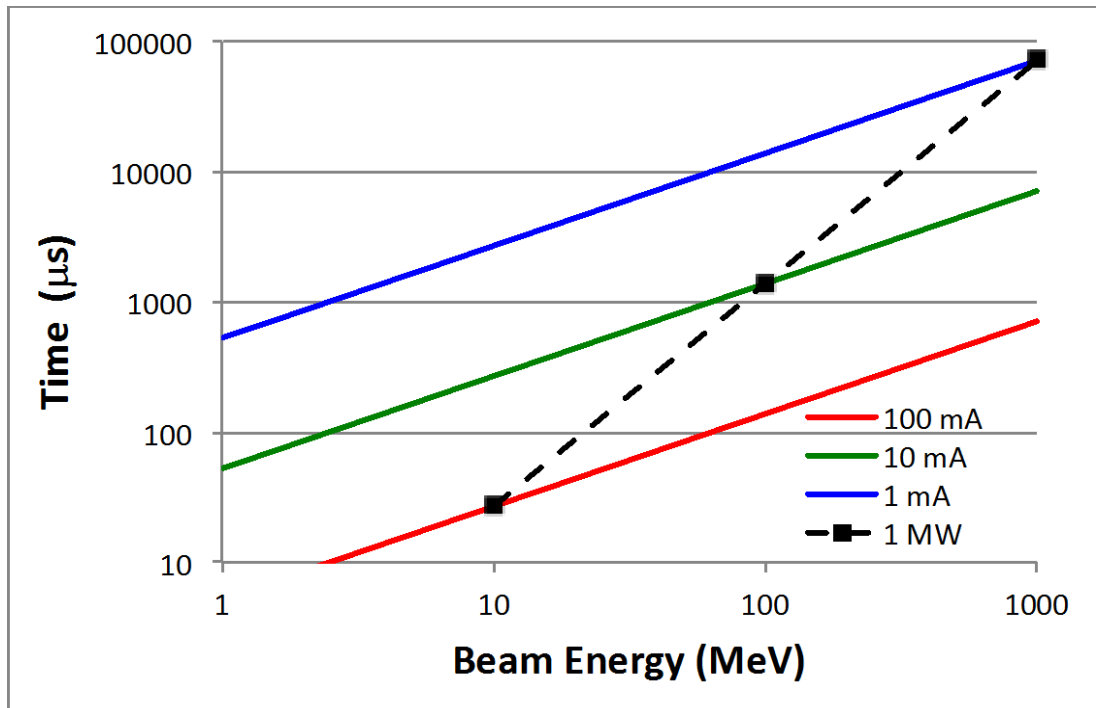

**Fig 1:** Time required for temperature of Cu to increase by 100°C when subjected to a 1 mm² proton beam of varying average current and energy. The dashed line indicates the 1 MW beam contour.

Beyond catastrophic melting of the equipment, a major concern for high-power hadron accelerators is keeping beam loss below the level at which hands-on maintenance becomes problematic. As mentioned previously, while there is no absolute cut-off for this beam loss level, a general rule of thumb is 1 W/m beam loss. To put this level of beam loss in perspective, the 1 MW beam power contour is shown as a dashed line in Fig. 1. At this beam power, fractional loss should be maintained below $10^{-6}$/m. However, it takes a considerable period (e.g. days) of beam loss at this level to create long-lasting residual activation that inhibits maintenance. Detecting modest increases in beam loss (e.g. tens of percent) over long periods (e.g. days) can indicate an emerging equipment issue, without seriously jeopardizing machine health.

Because the build-up time for residual activation is so long, this provides some relief in the beam loss monitoring systems that protect against this hazard. This is a good circumstance, as beam loss detection for fractional loss at the $10^{-6}$ level is often noisy. For example, with loss monitors near accelerating structures (e.g., a linac), there is often background X-ray generation that obscures the actual beam loss signal. A common practice for 'slow' protection against low levels of beam loss is to time average the loss signals to enhance the signal-to-noise ratio. Careful monitoring of the slow beam loss signals and correlating these with other equipment signals can shed light on emerging equipment issues. Some examples of this are shown later.

## 2.2 Slow protection

Machine protection systems will inhibit the beam if an unsafe situation exists, even before one attempts to start high-power operation. A primary example of this sort of protection is checking that all the vacuum system valves where the beam will be transported are in the open position. Other examples include ensuring that interceptive beam diagnostic devices (for example, wire scanners) are not inserted in any area in which high-power beams may be directed. These are straightforward equipment protection methods against catastrophic equipment damage, and are covered in other lectures in this series.

# 3 Measuring equipment

## 3.1 Electrical measurements

Particle accelerators use magnets as a primary means to guide and focus the particles. Typically, minimum stability control requirements on the magnetic fields are $\sim 10^{-3}$ for single-pass systems (linacs) and $\sim 10^{-4}$ for multipass systems (rings). Proper field levels are typically set up using beam methods. It is important to maintain the field levels at the desired values for long periods, after the set-up has been completed. Most magnets in accelerators are electromagnetic, and consist of multiturn windings to create the magnetic field. Ignoring hysteresis effects, field stability is controlled by maintaining a constant current through the circuit.

It is relatively straightforward to measure the current in a power supply, and appropriate control capability is specified when the power supplies are ordered. It is a simple matter to adopt software that monitors the power supply output to ensure that the current is at the appropriate set-point. Figure 2 shows an example of a normal quadrupole power supply current fluctuating with time, for a d.c. application.

Large accelerator facilities have large numbers of magnets to monitor, and it is standard practice to have automated software applications to 'snapshot' magnet parameters (e.g. current set-points and read-back values) when the machine is set up and operating well. These applications can also monitor these levels and report variations in live values relative to the 'golden' snapshot.

It is possible for the current read-back to be acceptable, but for there still to be a problem in maintaining the desired magnetic field. This can happen if some of the current is accidentally shunted around the desired current path, for example, if there is a turn-to-turn short in a multiturn magnet, or if there is a partial short to ground, as illustrated schematically in Fig. 3(a) (Fig. 3(b) shows an example cause of a ground fault interrupt). The latter example is referred to as a ground fault, and power supplies typically have a protective component called a ground fault interrupt to prevent damage of the cable or power supplies. However, it may be possible for a small amount of current to short-circuit the desired path through the magnet, and still be within the acceptable range of the ground fault interrupt comparator.

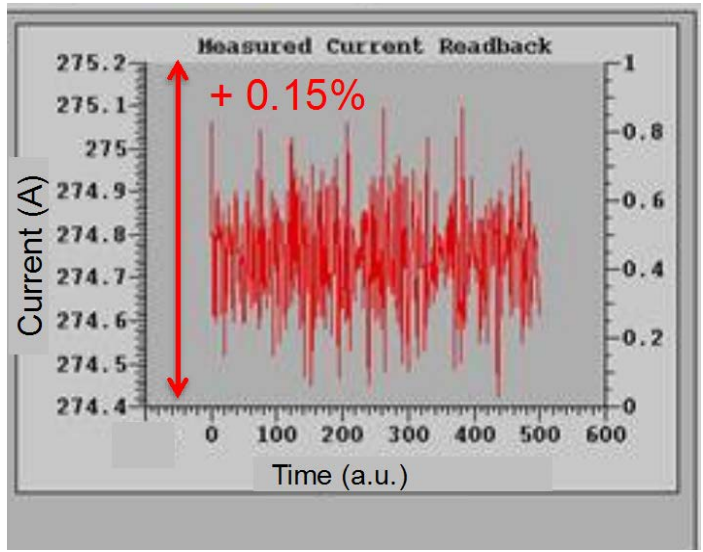

**Fig. 2:** Sample current fluctuation for a typical quadrupole power supply in a linac

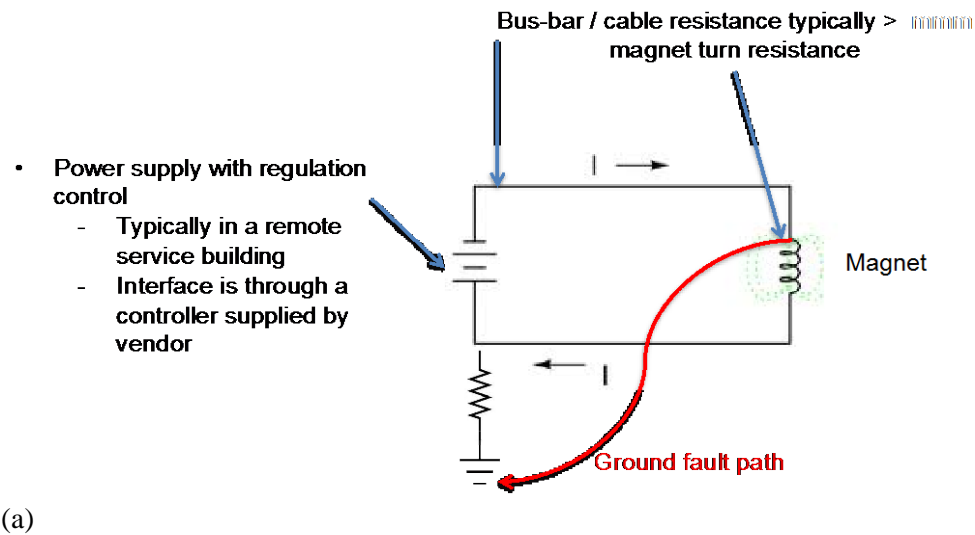

(a)

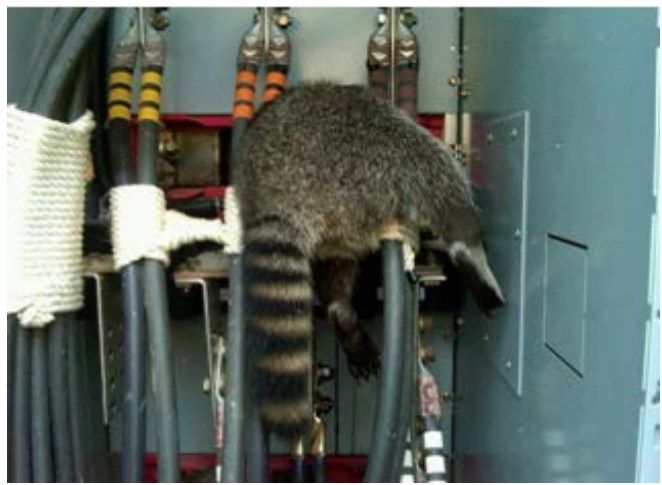

(b)

**Fig 3:** (a) Schematic of a ground fault that shunts part of the current around the desired path through a magnet. (b) Example cause of a ground fault interrupt.

For particle accelerators, the magnet power supplies are typically of the current-controlled type. A particular current is specified (corresponding to the desired magnetic field level), and the voltage is adjusted to produce the specified current. It is also possible to measure and monitor the voltage required to maintain the specified current. For d.c. magnets, the voltage is ideally constant; for cycled accelerators, the voltage can be quite complicated, but it should still follow the same pattern each cycle. A partial turn-to-turn short will slightly change the resistance across the magnet and affect the voltage required for the power supply to maintain the same current. However, typically in large accelerators, the resistance change is small compared with the overall circuit resistance (or impedance), which includes the effects from cables to and from the tunnel, and multiple magnets driven by the same power supply. The voltage change caused by a turn-to-turn short may not be detectable. Also, magnet power supplies are often not required to have tight tolerance on voltage read-back. Figure 4 shows a typical voltage read-back for two quadrupole power supplies in the SNS superconducting linac section. There is a large level of noise (>2%), which obviates the possibility of detecting small changes in the magnet coil resistance.

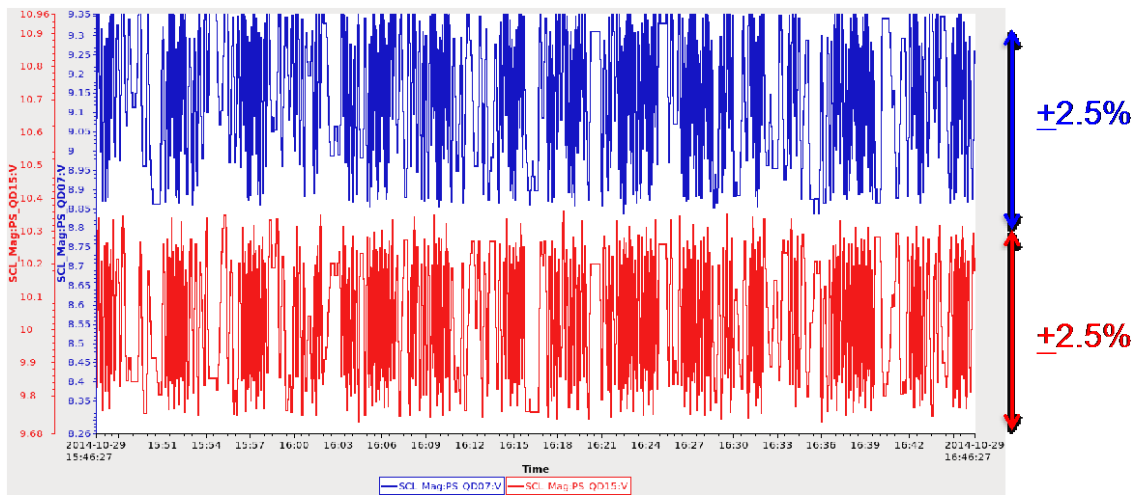

**Fig 4:** Time chart of the voltage read-back on two quadrupole power supplies (top and bottom plots) over ≈1 hour, indicating rather large noise fluctuations.

Although it may be difficult to detect equipment issues by voltage monitoring, as will be discussed in Section 4.1, it is possible to measure small changes in the applied magnetic field by monitoring changes in the beam trajectory.

### 3.2 Pulsed magnets

As mentioned already, cycled accelerators have repeatable patterns that magnets follow each cycle. While these may be complicated, there is typically a pattern goal for the magnet current over each cycle. Figure 5 indicates a dipole magnet current over one cycle. In this case, it is for an injection kicker in the SNS ring, and occurs over a period of a few milliseconds. Note that there are two curves: a target waveform (red), and the actual measured current waveform (blue). It is possible to compare the two waveforms electronically and report an error or alarm if the difference exceeds a pre-set value. This is the typical procedure for pulsed systems.

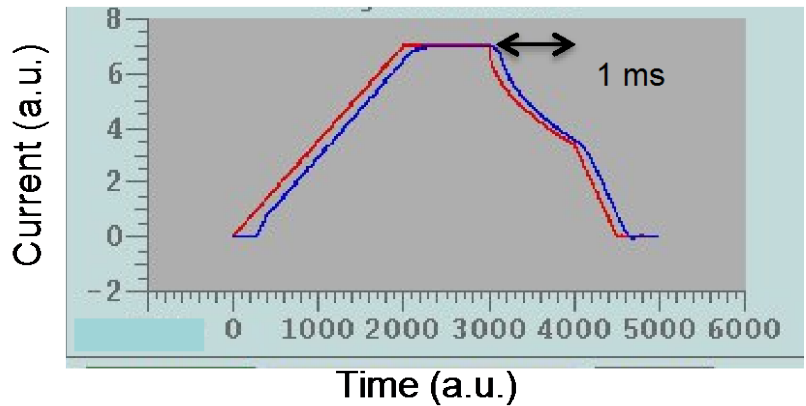

**Fig. 5:** Current waveforms for a pulsed dipole magnet, with a pre-set target waveform (red, initially leading curve) and actual read-back waveform (blue, initially trailing curve).

Another example of a pulsed magnet is a fast kicker system. In this case, magnets reach full field in a fraction of a turn within a ring (≈200–300 ns). The fields rise during a gap in the beam, so the precise waveform during the field rise is not important. The more critical issue in this case is the timing of the kicker firing, as premature or late firing results in the kicker affecting the beam, which is not in the gap. Figure 6 indicates a waveform display of a kicker in the SNS ring. There are actually ~6000 waveforms displayed on the plot, but they fall into two families, and appear as only two waveforms. The appearance of multiple traces indicates the initiation of some drift in the kicker firing, which can be a precursor of an emergent kicker problem. Persistent displays of this type are useful for illustrating a drift in time of a pulsed quantity.

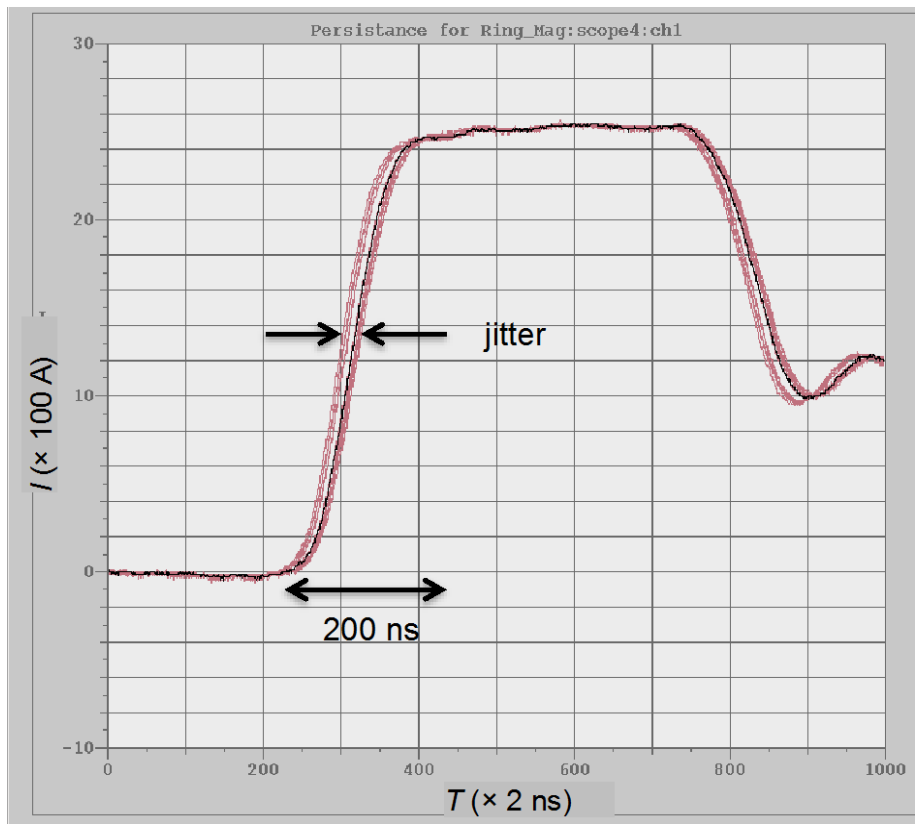

**Fig. 6:** A persistent display of a kicker waveform over many cycles, indicating a drift of the kicker firing

### 3.3 RF system monitoring

High-power RF systems are a major component of high-power accelerators. There are several linked subsystems within a pulsed high-power RF installation, as illustrated in Fig. 7. Power from the grid is converted from a.c. to high-voltage d.c. in a rectifier. For pulsed systems, the d.c. power is converted by some sort of a pulsed forming network to pulsed high-voltage d.c. waveforms, which are used to power an RF source. The RF power generated in the source is finally transported to the accelerating structure. A crucial part of the overall system is the low-level RF (LLRF) system, which coordinates the timing and amplitude control of the delivered RF power very precisely with the beam in the structure. Any variation or drift in the equipment of these components can cause beam loss. Some examples of identifying causes of beam loss due to RF equipment issues are shown here.

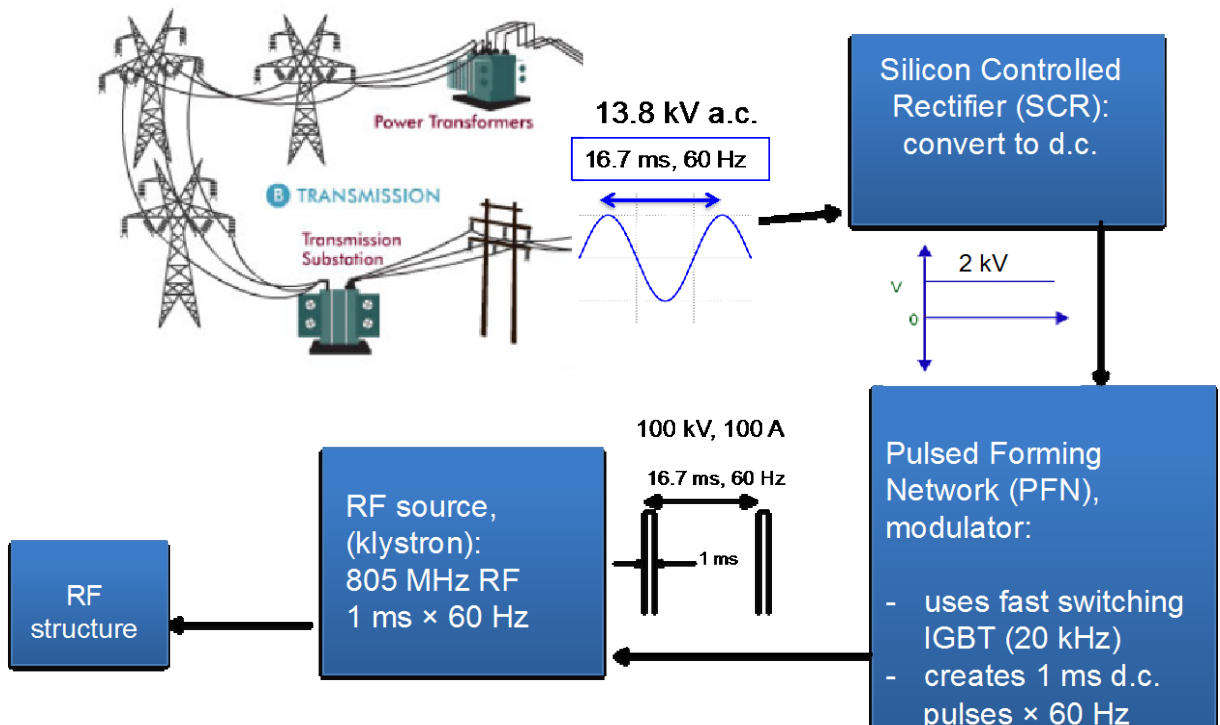

**Fig. 7:** Schematic of the components of a high-power RF installation

#### 3.3.1 LLRF issue

Figure 8 indicates a time history over ≈30 hours of: (1) the field gradient in a superconducting linac cavity (red), (2) the temperature of a downstream cavity beam-pipe (green), and (3) the beam arrival time in a downstream beam position monitor (blue). Near the middle of the time period, there is a sudden change in the beam-pipe temperature and arrival time of the beam, indicating a change in the beam acceleration. The beam loss monitors along the linac did not report a meaningful increase in beam loss, so the beam kept running. About 4 hours later, an operator noticed the change in the downstream beam arrival time and artificially increased the cavity gradient to restore the beam arrival time to the previous value. Subsequently, the beam-pipe temperature also returned to its previous level. In this case, the root cause of the observed changes was an electronic component failure in the LLRF system, resulting in a change in the field regulation, which the operator compensated for. The cavity gradient change was small (a few percent) so it was possible to continue running, but this was an indication of an equipment issue, which needed addressing. The elevated beam-pipe temperature was due to elevated beam loss, even though it was not detectable on loss monitors in this case.

This issue was diagnosed by correlating different signal changes along a timeline. Control systems have tools to perform this task, with both live streaming and archived data. This is a common technique for diagnosing previously unseen issues.

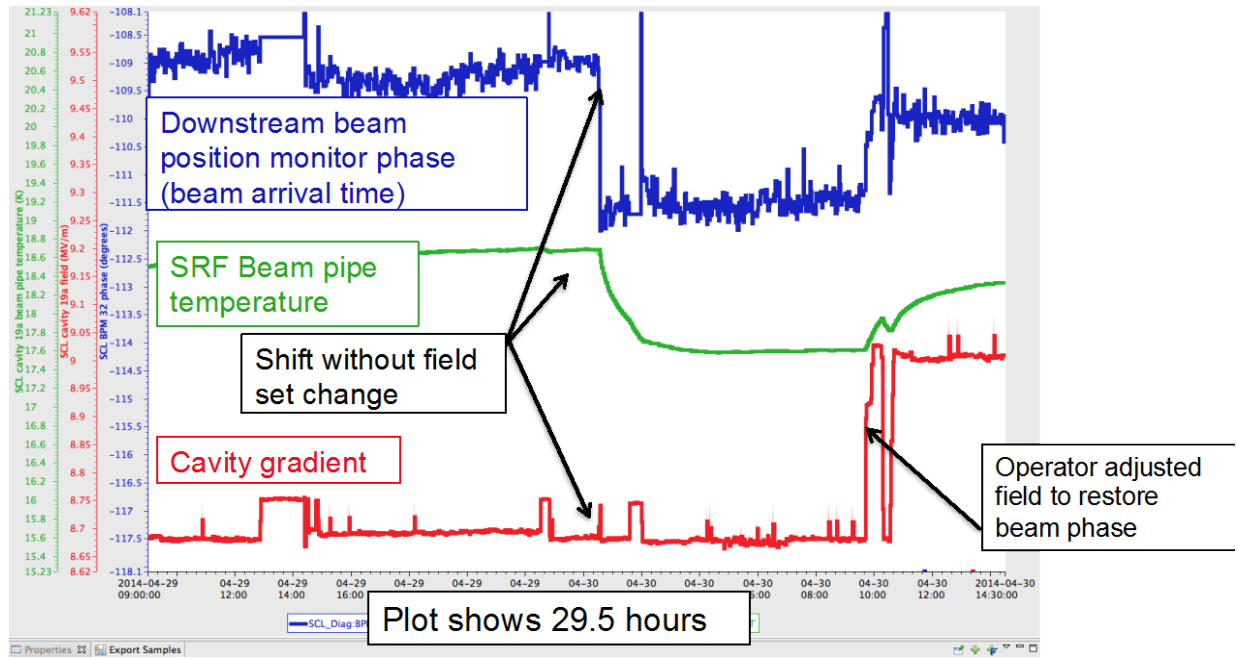

**Fig. 8:** Timeline indicating a shift in LLRF performance. The bottom curve is the cavity gradient, the middle curve is the beam pipe temperature, and the top curve is a downstream loss monitor signal. The entire time span shown is 29.5 hours.

### 3.3.2 Fast time-scale monitoring

Another technique for diagnosing RF system equipment issues is that of monitoring fast time-scale waveforms of the RF system. Figure 9(a) shows the LLRF amplitude waveform output for a 1 ms pulse in a copper cavity structure. Figure 9(b) shows a zoomed-in view of the amplitude axis. While the zoomed-out view indicates a fairly nice looking waveform, the zoomed-in image shows some amplitude noise at about the 0.5% level. This jitter is within the acceptable control margin, and is not, in itself, a concern. Analysis of the structure of the amplitude noise reveals a 20 kHz frequency component. It turns out that the pulse-forming network (PFN) uses 20 kHz solid-state switching technology to provide the high-voltage drive for the klystrons powering this cavity.

This amplitude fluctuation is an indicator of the PFN health, and can be monitored. If this ripple increases, it could become a source of beam loss, so maintenance or adjustment of the PFN unit can be identified from the downstream LLRF measurement. It is useful to have automated systems to monitor the quality of the RF waveforms and report cavities that exceed a permissible threshold of amplitude or phase variation.

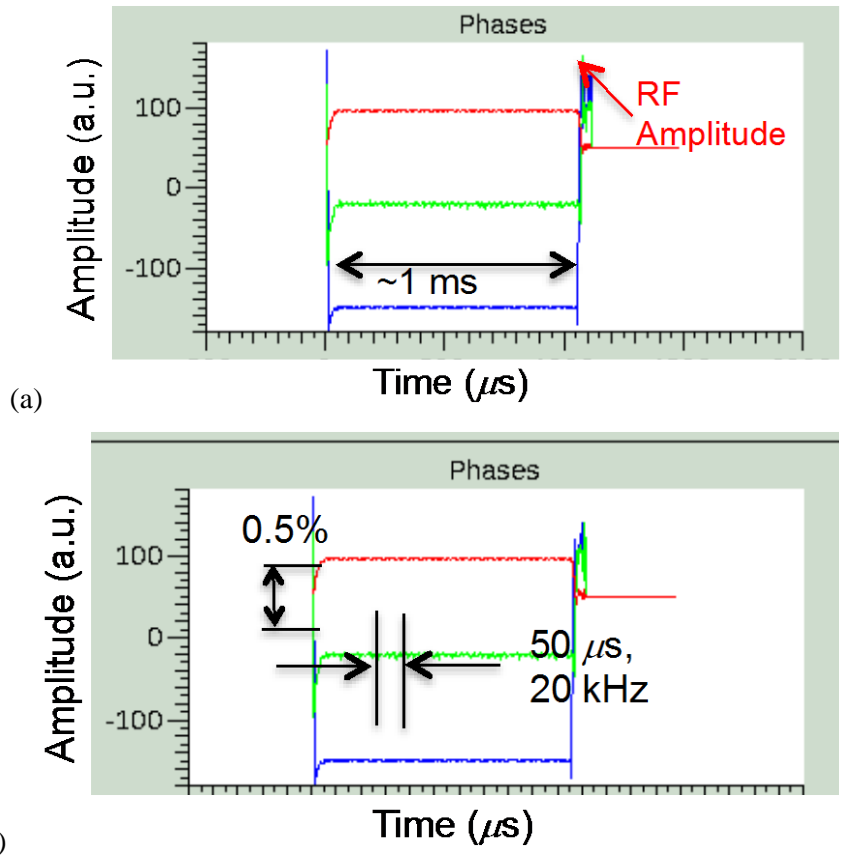

**Fig. 9:** (a) A waveform display of the RF amplitude (top curve in (a)) of a copper structure cavity. (b) Enlargement of the same trace, revealing 20 kHz noise.

### 3.3.3   Line-sync issues

High-power accelerators are large electrical power consumers. Historically, pulsed accelerators were designed to run at harmonics of the electrical grid frequency, to be able to 'ride along' at a constant phase of the electrical grid a.c. power cycle. Although electrical grids are referred to as operating at 60 Hz, or 50 Hz, there are constant slight variations in the electrical power generation frequency to match the demand load. Previous-generation pulsed accelerators would adjust the beam pulse timing to follow the grid frequency, so as to maintain a constant phase offset from the peak of the voltage cycle. With the advent of solid-state fast switching technology to provide PFN capabilities, modern electrical systems are gene rally not sensitive to when the beam pulse occurs relative to the grid power cycle. This enables accelerators to run at constant frequency, and even at a frequency that is not a harmonic of the grid. However, there can be issues if the grid a.c. frequency effects 'leak-through' to the beam in unanticipated ways. An example is shown in Fig. 10(a), which shows jitter in a kicker timing signal (black trace) and the variation of the beam trigger with grid line cycle peak (red trace). There is a weak correlation in these traces, which led to the investigation of the kicker timing unit. Indeed, a low-cost a.c.–d.c. transformer was found to be performing below its specification (with an unacceptable a.c. component leading through to the d.c. signal). Although the beam loss was acceptable, if the issue progressed, it would become an issue. Figure 10(b) shows the improved-stability timing signal after replacement of the faulty a.c.–d.c. conversion unit.

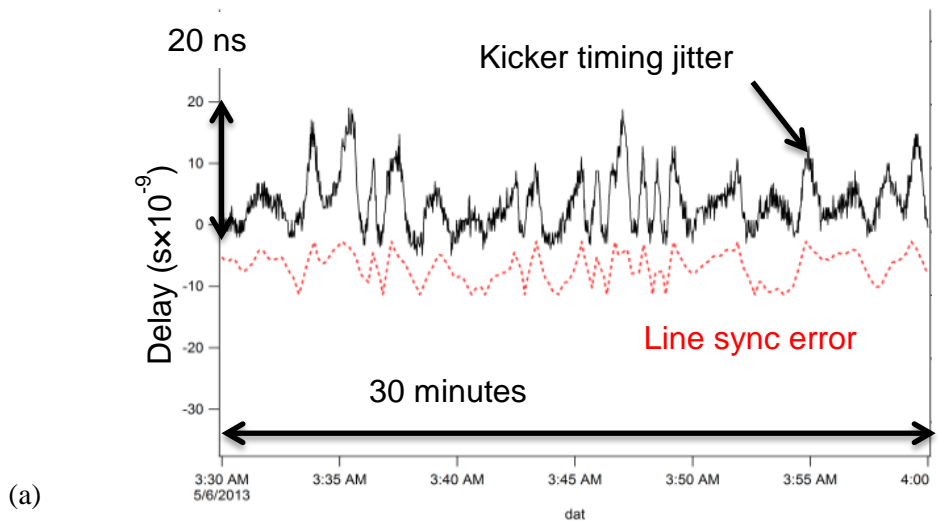

(a)

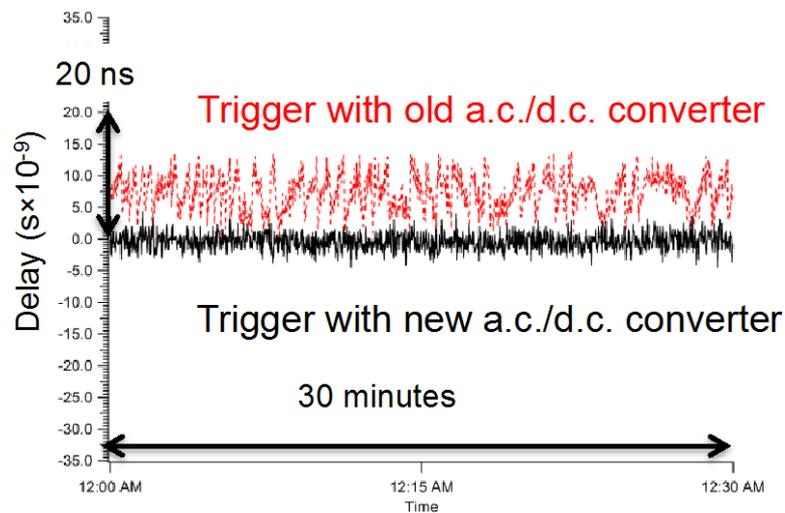

(b)

**Fig. 10:** (a) Jitter in the timing signal of a kicker magnet (black) with a correlation in the line-synch phase variation (red). (b) Jitter in the timing signal with the faulty a.c.–d.c. convertor (red) and with a replaced unit (black).

### 3.4 Vacuum

Most high-power accelerators require high vacuum to maintain low beam loss. Beam scattering and charge state changes (e.g. e-stripping) are examples of the loss mechanisms that cause beam loss [3]. Figure 11 indicates a clear response of downstream losses in a superconducting linac section, from a purposeful change in an upstream copper linac section vacuum level (caused by turning off vacuum pumps). While there is a clear dependence of loss on increasing vacuum levels, there is also another loss component. Monitoring the health of the vacuum systems is clearly an important part of preventing beam loss. Figure 12 shows the time history of the vacuum in a transport line of the SNS accelerator. There are fairly regular vacuum 'spikes' in the first half of the display, but these can be ignored. They happen regularly (e.g. related to beam trips and related loss). However, in the second half of the display, a sustained increase in vacuum baseline is observed, and is cause for implementing vacuum pump repair work. In this case, there was not a serious beam loss increase, but this is a precursor to more serious issues.

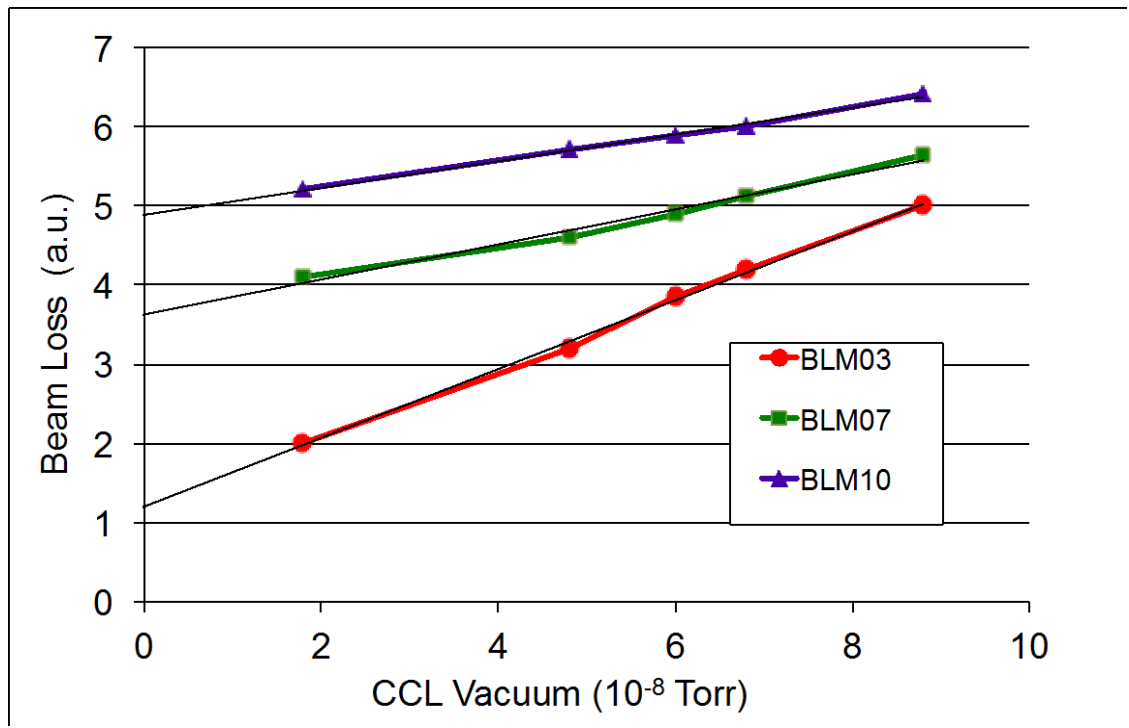

**Fig. 11:** Variation of downstream beam loss measurements with a purposeful variation of vacuum in an upstream region.

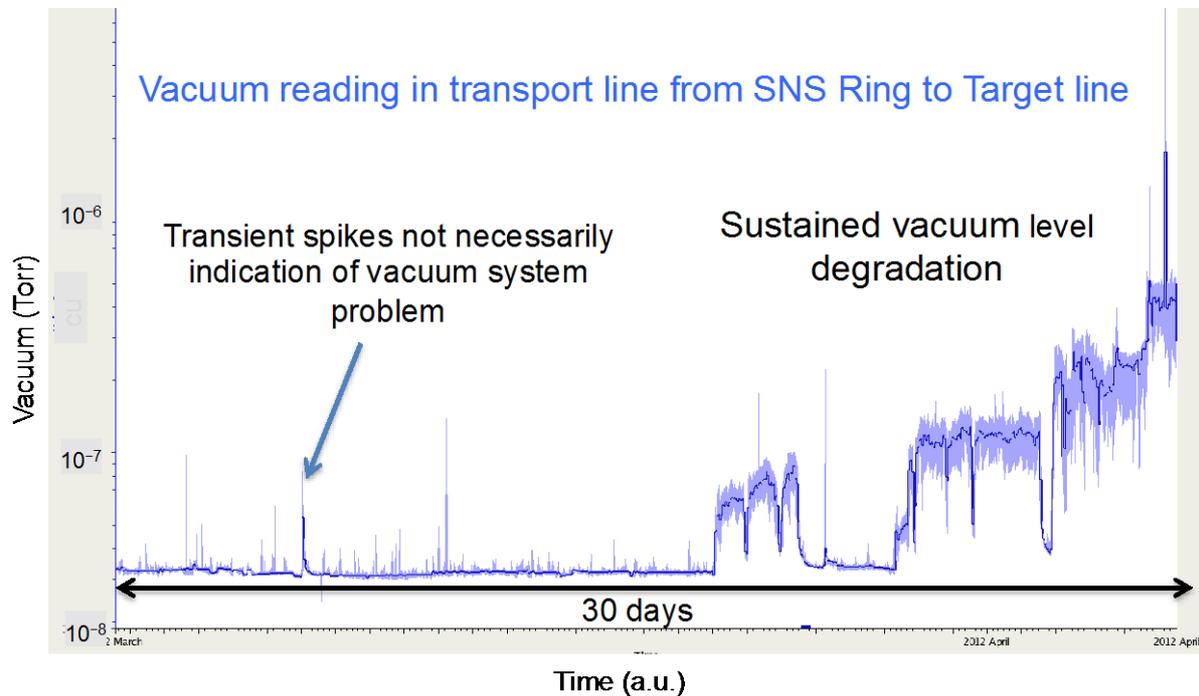

**Fig. 12:** One-month time history of the vacuum level in a transport line section, indicating a developing issue

### 3.5 Temperature monitoring

Measuring ambient air and equipment temperatures can be a useful tool in identifying equipment issues that cause beam loss. Much modern accelerator equipment is controlled by sensitive electronics, which may include temperature-sensitive components. For example, Ref. [4] describes how order of magnitude variations in the thermal stability of the LLRF analogue front end board stability were measured, dependent on the particular electrical components used on the board. Figure 13 shows the RF field amplitude stability over several hours for two cavities with similar LLRF electronics. The unit in a rack with temperature control stabilization shows an acceptable drift in the field level, whereas the unit without temperature control shows an unacceptable drift. Temperature control stabilization in the racks was instituted to alleviate the drift, as indicated in Fig. 13. The better solution is to use electronic components that are not temperature sensitive, but this is not always possible.

Another temperature sensitivity of concern is the change in cable lengths with temperature. For some critical applications, such as RF reference lines, steps are taken to minimize this effect (temperature control of the reference line). But it is prohibitively expensive to control the temperature of all cables, and some beam instrumentation cables and LLRF cables can be many tens of metres long, and subject to slight length changes, which can cause changes of a few degrees Fahrenheit in the RF phase control for high frequency (~GHz) systems. As an example of this effect, Fig. 14 shows the SNS klystron gallery building temperature (measured at several locations) and a downstream linac beam loss, measured over about 10 days. For the first week of the display, a new heating ventilation and air conditioning (HVAC) operational mode was attempted, which resulted in continuous temperature fluctuations of ~1°F throughout the building (this is the building that houses the RF equipment). Finally, the old control mode was re-established, and both building air temperature fluctuations and unexplained beam loss fluctuations disappeared.

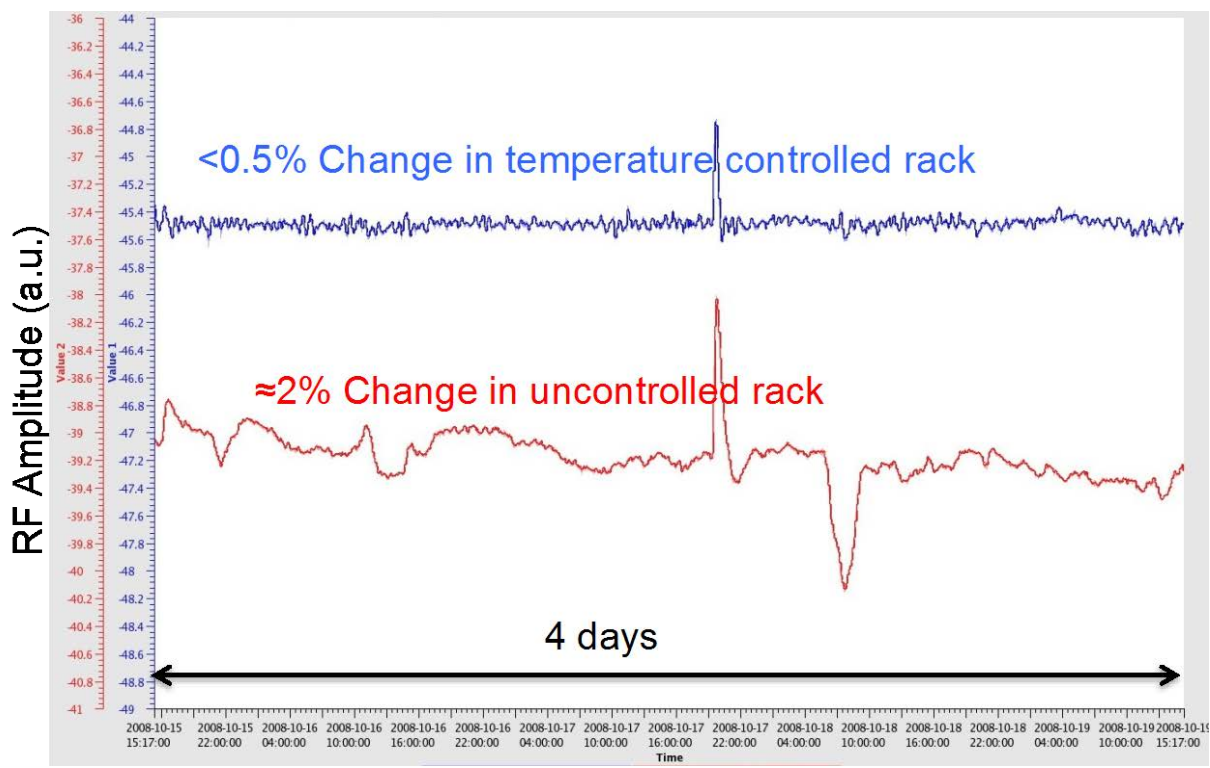

**Fig. 13:** Drifts in the RF field amplitude in a cavity with temperature-controlled LLRF electronics (blue) and in a cavity with similar LLRF electronics in a rack without temperature control (red).

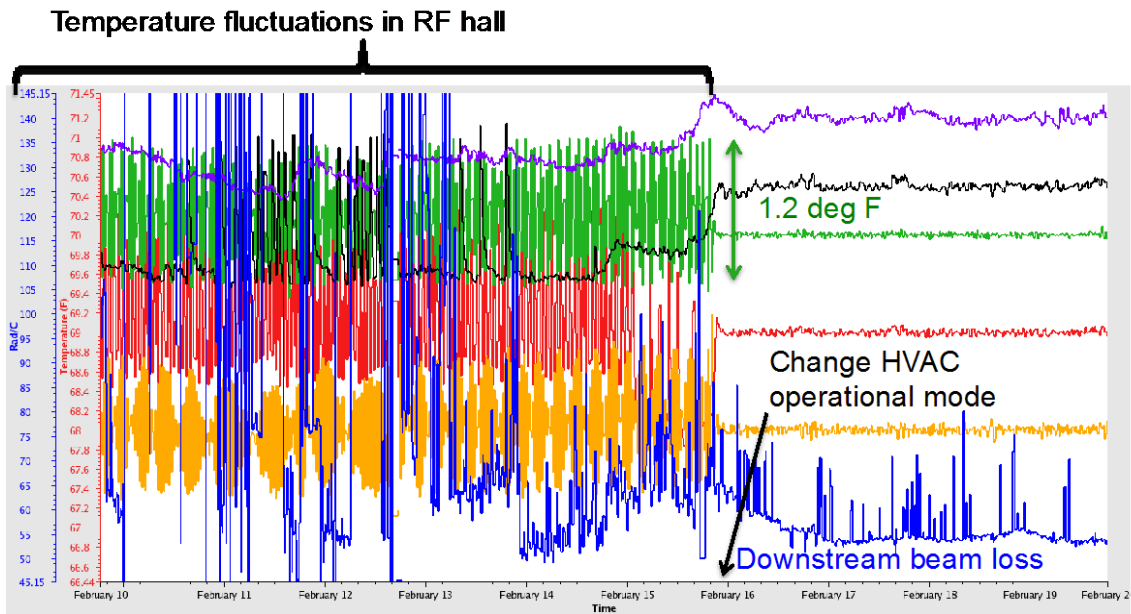

**Fig. 14:** Ten day period during which a new HVAC control mode was attempted, resulting in ~1°F fluctuations within the linac RF gallery [green (third curve from top), red (third curve from bottom), and orange (second curve from bottom) traces], causing unstable beam loss in the linac (bottom blue) trace.

Another example of a subtle temperature change affecting beam parameters is shown in Fig. 15, which shows parameter variations over ≈3 days. In this case, unexpected erratic behaviour was observed in the beam injection area of the SNS accumulator ring (green curve labelled 'beam missing foil'). An operator noticed a correlation of this parameter with the variation in amplitude of the linac RF reference signal (which should be constant). The erratic behaviour only occurred at night, when the building, which houses the reference line signal generator, was cooler. Replacing the reference line signal generator solved the issue, as it had developed a temperature sensitivity. This is another example of how searching for correlations helped identify the source of the equipment problem. The associated beam issues were never severe enough to stop beam operation in this case.

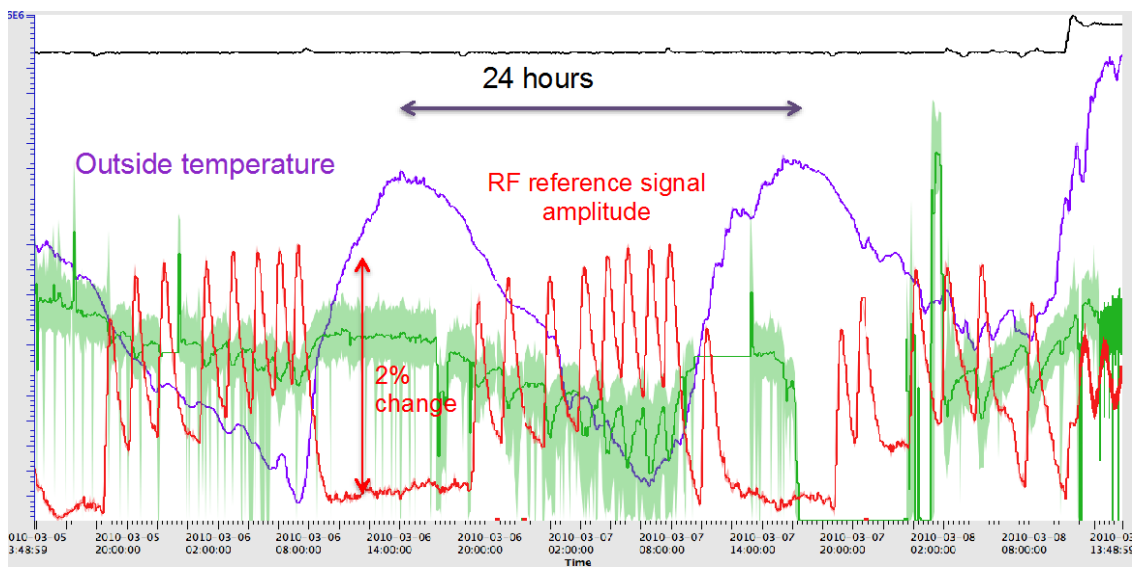

**Fig. 15:** Diurnal temperature variation leading to unexpected behaviour in RF reference line control and a related change in beam parameters.

Finally, we should note that it is often useful to monitor equipment temperature directly, even if direct temperature sensors are not available. Infrared imaging is a useful technique to provide highly localized temperature information. An example is shown in Fig. 16. This technique is useful for periodic monitoring of circuit breakers and magnet cable connections, as loose connections generate heat, which leads to equipment degradation or failure.

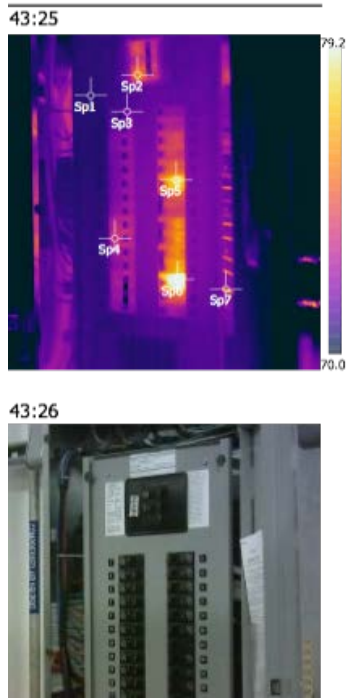

**Fig. 16:** Thermal (top) and visual (bottom) images of the same breaker panel. The thermal image shows hot spots (e.g. Sp2, Sp6, and Sp7) not seen in the visible spectrum image.

## 4    Beam measurements

The measurements described so far have focused on monitoring signals from equipment or buildings to understand emergent issues that may lead to beam loss. In addition, beam signals can be useful for identifying equipment issues. The beam has perhaps the most sensitive response to small changes in the equipment or external environment.

### 4.1    Transverse beam measurements

Measuring changes in the transverse beam position is the most direct method to detect issues with equipment that affect the transverse beam position (e.g. magnets and RF devices for hadron beams). Figure 17(a) shows a typical beam trajectory along a linac. Ideally the trajectory lies perfectly along the axis; however, there are typically slight imperfections. This trajectory is somewhat chaotic, but acceptable. Figure 17(b) indicates the same beam trajectory, with a slight change to a steerer at the location indicated. It is not obvious from this image alone where the trajectory change originated. However, a plot of the difference between the two trajectories indicates a wave in the beam motion beginning where the steering changed. Orbit (trajectory) difference techniques are powerful methods of identifying changes in steering magnets and RF or quadrupole elements if the beam is even slightly off axis. Sometimes, the settings of the magnet or RF may not have changed. For example, if a slight turn-to-turn short is developing, the applied magnetic field can change slightly, even though the magnet current settings have not been changed. Also, if the LLRF control is not working properly (e.g. as shown in Fig. 8), a trajectory change may be observable.

The key to applying this method is to take a snapshot of the beam transverse trajectory along the accelerator during the set-up period, when the quality is known to be in a good state. Software can be provided to highlight changes in the saved ('golden') and live beam trajectories along the accelerator.

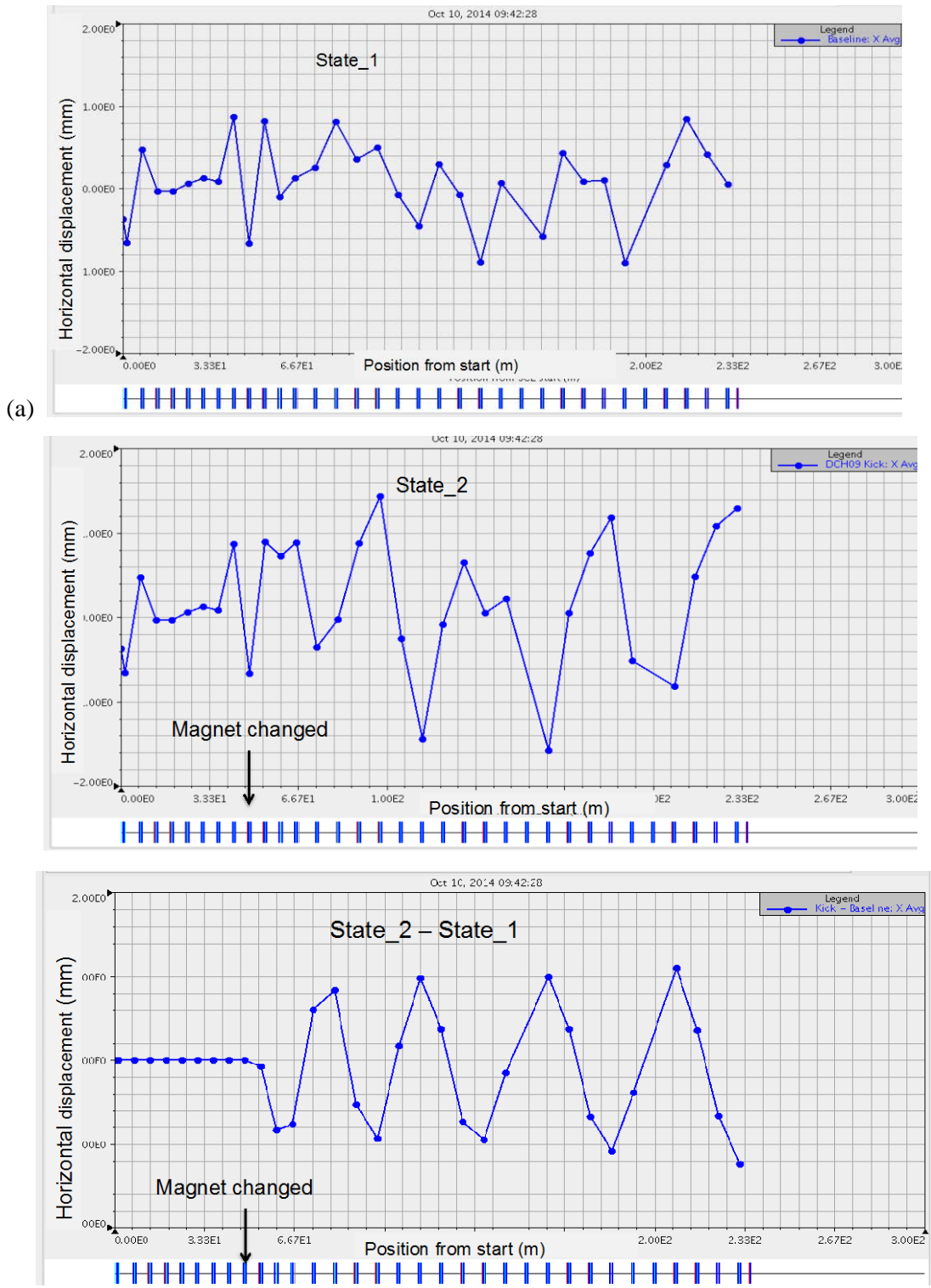

(a)

(b)

(c)

**Fig. 17:** (a) Original beam trajectory along the horizontal axis. (b) Trajectory after a change in the steering at the indicated location. (c) Difference between the trajectories of (b) and (a). (Magnets are displayed synoptically below the x-axis).

## 4.2 Longitudinal beam measurements

The method described in Section 4.1 can also be applied in the longitudinal direction. In the longitudinal plane, the longitudinal 'position' is typically recorded as the beam arrival time relative to a reference RF signal, often in units of 'degrees' of the reference RF signal. The measuring device may be a beam position monitor (if properly designed for this capability), or an RF resonator of some sort. Figure 18(a) displays the change in beam arrival time along the SNS linac from the saved beam set-up values. A wave is evident, starting from the beginning (the magnitude of the wave changes along the linac in this case, because the units of the arrival time change, and are not corrected here). This provided information as to where the problem originated, and an operator adjusted the phase of the first cavity in the linac, resulting in the difference plot shown in Fig. 18(b), which shows that the beam trajectory is much closer to the set-up conditions. This enabled continued beam running until a maintenance day, when a poor LLRF cable contact was discovered at the first cavity.

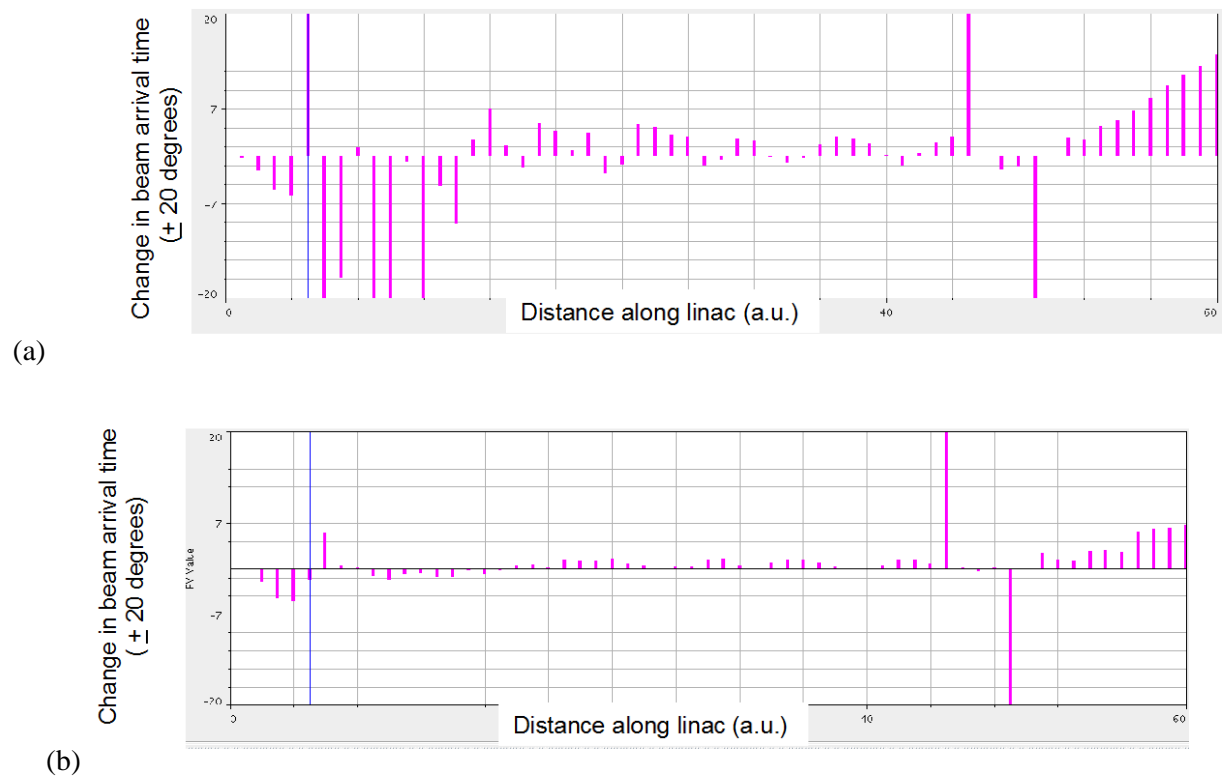

(a)

(b)

**Fig. 18:** (a) Change in beam arrival time along the SNS linac from beam set-up conditions. (b) Same plot as (a), after adjusting the phase of an RF structure at the start of the linac.

## 5 Summary

Identifying emergent equipment issues before a problem results in unacceptable beam loss is a challenging task. For high-power accelerators, beam loss becomes intolerable at quite small fractional levels, but it is possible to operate with some small level of beam loss increase. The challenge lies in identifying the equipment that causes the change in beam loss. Different equipment properties have been identified for monitoring, and example techniques for monitoring these quantities are shown. A key element in these techniques involves identifying correlations between changes in beam parameters and equipment parameters to identify the culprit driver for the onset of beam loss increase. Finally, using direct beam measurements to localize the source of equipment issues was described.

## Acknowledgements

ORNL is managed by UT-Battelle, LLC, under contract DE-AC05-00OR22725 for the US Department of Energy.

Notice: This manuscript has been authored by UT-Battelle, LLC, under Contract No. DE-AC05-00OR22725 with the US Department of Energy. The US Government retains, and the publisher, by accepting the article for publication, acknowledges, that the US Government retains a non-exclusive, paid-up, irrevocable, world-wide license to publish or reproduce the published form of this manuscript, or allow others to do so, for US Government purposes.